\documentclass[twocolumn,prl,aps,showpacs]{revtex4}
\usepackage{amssymb}

\usepackage{amsfonts}
\usepackage{amsmath}
\usepackage{epsfig}

\begin{document}

\title{Attosecond Interference Induced by Coulomb-Field-Driven Transverse Backward-Scattering Electron Wave-Packets}

\author{Xiaohong Song}

\affiliation{Department of Physics, College of Science, Shantou
University, Shantou, Guangdong 515063, China}

\author{Peng Liu}

\affiliation{Department of Physics, College of Science, Shantou
University, Shantou, Guangdong 515063, China}

\author{Cheng Lin}

\affiliation{Department of Physics, College of Science, Shantou
University, Shantou, Guangdong 515063, China}

\author{Zhihao Sheng}

\affiliation{Department of Physics, College of Science, Shantou
University, Shantou, Guangdong 515063, China}

\author{Xianhuan Yu}

\affiliation{Department of Physics, College of Science, Shantou
University, Shantou, Guangdong 515063, China}

\author{Weifeng Yang}
\email{wfyang@stu.edu.cn}

\affiliation{Department of Physics, College of Science, Shantou
University, Shantou, Guangdong 515063, China}

\author{Shilin Hu}

\affiliation{HEDPS, Center for Applied Physics and Technology,
Collaborative Innovation Center of IFSA, Peking University,
Beijing 100084, China}

\affiliation{Institute of Applied Physics and Computational
Mathematics, P. O. Box 8009, Beijing 100088, China}

\author{Jing Chen}
\email{chen_jing@iapcm.ac.cn}

\affiliation{HEDPS, Center for Applied Physics and Technology,
Collaborative Innovation Center of IFSA, Peking University,
Beijing 100084, China}

\affiliation{Institute of Applied Physics and Computational
Mathematics, P. O. Box 8009, Beijing 100088, China}

\author{SongPo Xu}
\affiliation{State Key Laboratory of Magnetic Resonance and Atomic
and Molecular Physics, Wuhan Institute of Physics and Mathematics,
Chinese Academy of Sciences, Wuhan 430071, China}

\author{YongJu Chen}
\affiliation{State Key Laboratory of Magnetic Resonance and Atomic
and Molecular Physics, Wuhan Institute of Physics and Mathematics,
Chinese Academy of Sciences, Wuhan 430071, China}

\author{Wei Quan}
\affiliation{State Key Laboratory of Magnetic Resonance and Atomic
and Molecular Physics, Wuhan Institute of Physics and Mathematics,
Chinese Academy of Sciences, Wuhan 430071, China}

\author{XiaoJun Liu}
\email{xjliu@wipm.ac.cn} \affiliation{State Key Laboratory of
Magnetic Resonance and Atomic and Molecular Physics, Wuhan
Institute of Physics and Mathematics, Chinese Academy of Sciences,
Wuhan 430071, China}

\date{\today}

\begin{abstract}

A novel and universal interference structure is found in the
photoelectron momentum distribution of atoms in intense infrared
laser field. Theoretical analysis shows that this structure can be
attributed to a new form of Coulomb-field-driven
backward-scattering of photoelectrons in the direction perpendicular
to the laser field, in contrast to the conventional rescattering
along the laser polarization direction. This transverse
backward-scattering process is closely related to a family of
photoelectrons initially ionized within a time interval of less
than 200 attosecond around the crest of the laser electric field.
Those electrons, acquiring near-zero return energy in the laser
field, will be pulled back solely by the ionic Coulomb field and
backscattered in the transverse direction. Moreover, this
rescattering process mainly occurs at the first or the second
return times, giving rise to different phases of the
photoelectrons. The interference between these photoelectrons
leads to unique curved interference fringes which are observable
for most current intense field experiments, opening a new way to record the
electron dynamics in atoms and molecules on a time scale much
shorter than an optical cycle.

\end{abstract}


\maketitle

In the ionization process of atoms in intense laser field, the
electron wave packet (EWP) may follow different paths from its
bound state to the continuum in the combined laser and Coulomb
fields. The interference between the EWPs might create richly
structured patterns in the final photoelectron distribution, which
inherently encode the temporal and spatial information of the ions
and electrons. For example, a holographic interference structure
was recently observed in the photoelectron momentum distribution
(PMD) of metastable xenon atom ionized by a 7 $\mu$m free-electron
laser pulse \cite{Huismans7}. This interference structure was
explained as interference between the direct and the laser-driven
forward-scattered EWPs generated within the same quarter-cycle of
the laser pulse, providing an efficient way in exploring the
structure and the dynamics of the atoms and molecules with
attosecond temporal and angstrom spatial resolution. Thereafter,
the photoelectron interference structure has been extensively
investigated for a broad range of laser parameters covering
tunneling to multiphoton ionization regimes, however, a full
understanding of its underlying physics has not yet been achieved
\cite{Hickstein8,Meckel9,Huismans10,Marchenko11,Yang12,Bian12,Bian13}.

Note that for the formation of this specific holographic
interference structure, the reference wave, i.e., the direct
electron, upon ionization, was assumed to be very weakly affected
by the ionic Coulomb field, while the rescattered one, experienced
strong Coulomb focusing and passed close to the ion, was
considered as the signal wave. Moreover, previous studies have
suggested that the Coulomb field plays a negligible role in the
holographic interference patterns
\cite{Huismans7,Huismans10,Hickstein8}. On the other hand, it has
been generally accepted that the ionic Coulomb field plays a
pivotal role in the photoelectron dynamics, e.g., giving rise to
an unexpected ``low-energy" structure in the photoelectron energy
spectrum
\cite{Blaga2009NatPhys,Quan2009PRL,Wu2012PRL,liu10,yan10,rost12,Guo2013,Becker15}
and a clear minimum at zero in the electron momentum distribution
along the laser polarization direction
\cite{Chen02,Moshammer2003PRL}. An open question thus concerns if
the ionic Coulomb field would find its fingerprints in more
general interference patterns, if not in the specific holographic
interference, and more importantly, under which circumstance and
to which extent the ionic Coulomb field would play a role in the
interference pattern.


In this letter, we show that the ionic Coulomb potential can leave
a significant imprint on the interference structure in the
photoelectron momentum spectrum of atoms, which is demonstrated
experimentally by a novel and universal curved interference
pattern in the PMD. Using a recently developed generalized
quantum-trajectory Monte Carlo (GQTMC) method, we clarify that
this new structure can be attributed to a Coulomb-field-driven
transverse backward-scattering process. In contrast to
conventional rescattering which happens in the laser polarization
direction, when electrons emitted around the peaks of the laser
electric field come back to the core with near-zero drift energy
in the laser polarization direction, they will be pulled back solely by the
Coulomb potential and backward scattered upon the core in the direction of perpendicular to laser polarization axis. The interference among the
Coulomb-field-driven transverse backward-scattering electrons,
initially emitted near the crest of the oscillating electric
field, can induce a distinct interference structure, which can be
well distinguished from other interference structures, e.g., the
well-documented holographic interference structure.

The experiments have been performed with cold target recoil-ion
momentum spectroscopy (COLTRIMS) \cite{Ullrich2011,Jahnke2004}.
The laser pulse was generated by a commercial Ti: Sapphire
femtosecond laser system (FEMTOPOWER, Femtolasers Produktions
GmbH) with a center wavelength of 800 nm, a pulse duration of 30
fs and a repetition rate of 5 kHz. The pulse energy was controlled
by means of an achromatic half wave plate followed by a polarizer.
The laser beam was directed and focused into a supersonic Ar gas
jet inside the COLTRIMS vacuum chamber. The photoelectrons and
Ar$^{+}$ ions created in the interaction region were accelerated
by a uniform weak electric field (3.8 V/cm) towards two
position-sensitive Microchannel plate (MCP) detectors. A pair of
Helmholtz coils generated a weak uniform magnetic field (7.8
Gauss) to confine the electron movement perpendicular to the
electric field. The three dimensional vector momenta of
photoelectrons and Ar$^{+}$ were obtained from their times of
flight and the impact positions.

\begin{figure}[h]
\centering
\includegraphics[width=0.45\textwidth,height=0.25\textwidth,angle=0]{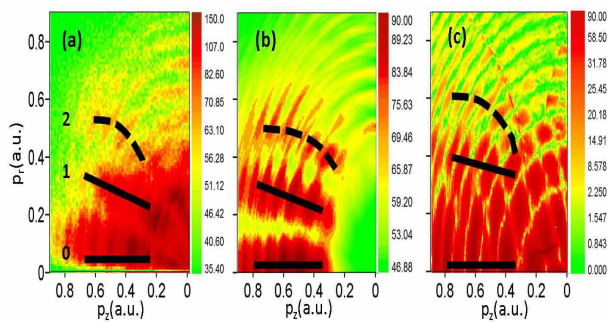}
\caption{(Color online) Experimental and simulated two-dimensional
photoelectron momentum spectra of Ar atom. (a) Experimental
result; (b) Focal-averaged TDSE simulation; (c) GQTMC simulation. Laser intensity
$I=1.7\times{10^{14}}$ W/cm$^{2}$, the wavelength $\lambda=800$
nm, and the pulse duration is 30 fs.}
\end{figure}

Fig. 1(a) shows the experimental PMD from Ar atom driven by an
800nm laser pulse with intensity of $1.7\times{10^{14}}$
W/cm$^{2}$. The laser pulse is linearly polarized
along the z axis. Holographic interference stripes (marked by
solid line) can be clearly seen \cite{Huismans7}. Moreover, an
additional interference fringes (marked by dash line) can also be
observed in Fig. 1(a). Unlike the holographic interference fringes
which are almost straight \cite{Bian13}, this interference fringe
clearly shows an arc shape. Calculation using the time-dependent
Schr\"{o}dinger equation (TDSE) \cite{Yang12} well reproduces the
experimental observation including both the conventional
holographic and curved stripes (see Fig. 1(b)). In the
simulation, focal-averaging over the laser intensities in the
focus is considered.

To reveal the underlying mechanism of this new interference
structure, we apply a generalized quantum-trajectory Monte Carlo
method to calculate the PMD which is depicted in Fig. 1(c). The
GQTMC method is based on the nonadiabatic ionization theory
\cite{Yudin3,Perelomov}, classical dynamics with combined laser
and Coulomb fields \cite{Brabec1996PRA,Hu14,Chen2000}, and the
Feynman's path integral approach \cite{Salieres2001,Li2014}.
Briefly, the evolution of the EWPs in the laser field is simulated
by launching randomly a set of electron trajectories with
different initial conditions. The weight of each electron
trajectory is given by
$w(\textbf{v}_{r0},t_{0})=\Gamma(t_{0})\Omega(\textbf{v}_{r0},t_{0})$.
Here $\Gamma(t_{0})$ is the ionization rate. Different from the
conventional QTMC model \cite{Li2014} where the
Ammosov-Delone-Krainov (ADK) ionization rate is used, our GQTMC
method is based on the nonadiabatic ionization theory
\cite{Yudin3,Perelomov} so that it can be applied to a quite large
Keldysh parameter range. The details and the comparison among the
TDSE, QTMC and GQTMC methods will be described elsewhere
\cite{Song}. The instantaneous ionization rate is given as
\cite{Yudin3,Perelomov}

\begin{equation}
\Gamma(t)=N(t)\exp(-\frac{E_{0}^{2}f^{2}(t)}{\omega^{3}}\Phi(\gamma(t),\theta(t))).
\end{equation}
Here, $E_{0}f(t)$ and $\theta(t)$ are the envelope and the phase
of the laser electric field, respectively. The pre-exponential
factor is
\begin{eqnarray}
&N(t)=A_{n^{*},l^{*}}B_{l,|m|}(\frac{3\kappa}{\gamma^{3}})^{\frac{1}{2}}CI_{p}(\frac{2(2I_{p})^{3/2}}{E(t)})^{2n^{*}-|m|-1} \nonumber \\
\label{eq:12}\\
&\kappa=\ln(\gamma+\sqrt{\gamma^{2}+1})-\frac{\gamma}{\sqrt{\gamma^{2}+1}}\nonumber
\end{eqnarray}
where $I_{p}$ the ionization potential, and the coefficient
$A_{n^{*},l^{*}}$ and $B_{l,|m|}$ coming from the radial and
angular part of the wave function are given by Eq.(2) in Ref.
\cite{Yudin3}. $C=(1+\gamma^{2})^{|m|/2+3/4}A_{m}(\omega,\gamma)$
is the Perelomov-Popov-Terent'ev correction to the quasistatic
limit $\gamma\ll1$ of the Coulomb pre-exponential factor with
$A_{m}$ given by Eqs. (55) and (56) in Ref. \cite{Perelomov}.
$\Omega(\textbf{v}_{r0},t_{0})=[\textbf{v}_{r0}\sqrt{2I_{p}}/|E(t_{0})|]\exp[\sqrt{2I_{p}}(\textbf{v}_{r0})^{2}/|E(t_{0})|]$
is the electrons' initial transverse velocity distributions. For
consistence, the coordinate of the tunnel exit is also modified to
be as
$z_{0}=\frac{2I_{p}}{E(t_{0})}(1+\sqrt{1+\gamma^{2}(t_{0})})^{-1}$
\cite{Perelomov}. After ionization, the classical motion of the
electrons in the combined laser and Coulomb fields is governed by
the Newtonian equations. The phase of the \emph{j}th electron
trajectory can be expressed as \cite{Salieres2001,Li2014}
\begin{equation}
 S_j (p,t_0
)=\int_{t_0}^{\infty}\{(\frac{v_p^2
(\tau)}{2}+I_p-\frac{Z_{eff}}{|r(\tau)|} )\}d\tau,
\end{equation}
where $p$ is the asymptotic momentum of the electron,
$Z_{eff}=\sqrt{2I_{p}}$ is the effective charge of the ion
potential. The EWPs will interfere with each other when they have
the same asymptotic momenta. Using a parallel algorithm, one
billion electron trajectories were calculated and added coherently
to obtain the PMD. The probability of each asymptotic momentum is
determined by
\begin{equation}
                |\Psi_p|^2=|\sum_k\sqrt{\Gamma(t_0,v_กอ^k )}\exp(-iS_k ) |^2.
\end{equation}

As shown in Fig. 1(c), the main features of the interference
fringes observed in Figs. 1(a) and 1(b) can be well reproduced by
the GQTMC simulation. To gain more insight into the origin
of the curved interference structure, we analyze all electron
trajectories contributing to the momentum spectrum with final
longitudinal momentum in the range ${0.3
a.u.}\leq\textbf{p}_{z}$. In this region, both
holographic fringes and curved interference fringes can be clearly
seen. Fig. 2(a) shows the distribution of the initial tunneling
phase and initial transverse momentum of the trajectories
contributing to this momentum range. It can be seen that, for the
momentum ranges we are analyzing, the initial conditions of
electrons within one laser cycle are separated into four areas
(denoted as area A-D in Fig. 2(a)). Figs. 2(b) and 2(d) show the
typical electron trajectories from areas B and A, respectively.
Obviously, these two areas correspond to different families of the
electron trajectories.

There are two types of electron trajectories in area B (see Fig.
2(b)). Electrons with small initial transverse momenta in this
area are forward scattered by the ionic potential in the direction
of the laser polarization (black lines). While the electrons with
large initial transverse momenta only revisit and pass by the core
at large distances without scattering, which are considered as the
direct electrons (red lines). To identify their
contributions to the total momentum spectrum, we then reconstruct
the final momentum distribution of electron trajectories only in
area B, as shown in Fig. 2 (c). In good agreement with pervious
strong-field approximation \cite{Huismans7} and classical
calculations \cite{Bian13}, the interference between these two
kinds of EWPs from area B yields the holographic interference
structures which is straight and radial. It reproduces the 0th and
1st fringes in the total photoelectron spectrum (see Fig. 1(a)).
Obviously, the electron trajectories that are initially launched
within area B do not lead to the curved interference pattern.

\begin{figure}[h]
\centering
\includegraphics[width=0.48\textwidth,height=0.36\textwidth,angle=0]{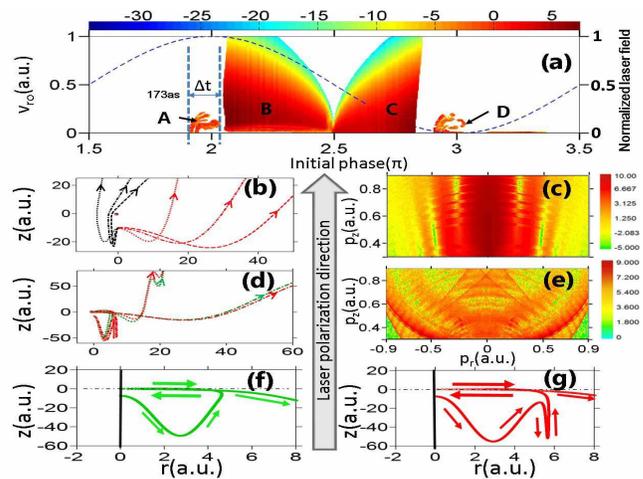}
\caption{(Color online) (a) Distributions of the initial
transverse velocities and the initial ionization phases for
$0.3 a.u.\leq\textbf{p}_{z}$ in Fig. 1(c). The color code denotes the weights of the electrons in area A-D. (b) typical trajectories of electrons in area B. (c) Reconstructed final
momentum distributions of electrons in area B. (d) typical
trajectories of electrons in area A. (e) Reconstructed final momentum distributions of electrons in area A.
A zoom in of electron trajectories scattered at the first and second return in (d) is shown in (f) and (g), respectively.}
\end{figure}

In Fig. 2 (d), it can be found that electron trajectories
initially launched in area A are quite different from those in
area B. As more clearly shown in Fig. 2(f), when the electron
comes back to around z=0, (i) it has near-zero returning velocity
along the laser polarization, so it almost stops at around z=0 in
the laser polarization i.e., $v_{z}\approx0$. (ii) it is the
Coulomb field that pulls the electron back along the line $z=0$ to
the core and further induces backward scattering in the direction
perpendicular to the laser polarization axis (also along the line
$z=0$). These two key points make this type of electron
trajectories quite special and distinguished in essence from
previous widely accepted rescattering process in which the
electrons are driven back mainly by the laser field and collide in
the laser polarization direction \cite{Faisal2009}. In this
Letter, we refer to this special rescattering as
``Coulomb-field-driven transverse backward-scattering". Moreover,
this Coulomb-field-driven transverse backward-scattering may occur
not only at the first time (see Fig. 2(f)) but also at the second
time (see Fig. 2(g)) when the electron returns to the core
\cite{note}. After scattering upon the core, the electrons may
possess the same momentum but apparent different phases, giving
rise to the interference fringes observed in Fig. 2(e). It is
worthwhile mentioning that the effect of the multiple return
trajectories in the interference structure in the PMD has been
reported in Ref. \cite{Hickstein8}. However, the structure
discussed there locates in small momentum region and, similar to
that proposed in Ref. \cite{Huismans7}, can be attributed to the
interference between undistorted and forward scattered electrons.

Obviously, the Coulomb potential plays a dominant role in the
evolution of this kind of electron trajectories. Without the
Coulomb potential, these electrons will only contribute to the
momentum map around $\textbf{p}_{z}\sim0$. Recently, it has been
reported that the interference between such direct ionization
electrons emitted at every electric field's extreme, which are
spaced by $T/2$, will result in a $2\hbar\omega$ separation of the
ATI rings for the perpendicular emission \cite{Korneev2012}.
However, once such electrons are driven back by the Coulomb
potential and further backward scattered in the transverse
direction, they may have large final momenta both in the
longitudinal and transverse directions.

Fig. 2(e) shows the reconstructed final momentum distribution of
electrons solely in area A. Most interestingly, the interference
among the electrons emitted from one single electric field
extremum but experienced Coulomb-field-driven transverse
backward-scattering at different return times can induce a novel
interference structure. These fringes show an arc pattern
different from the holographic interference fringes induced by
electron trajectories
in area B (see Fig. 2(c)) \cite{note2}. 
Clearly, it gives rise to the curved 2nd fringes in both the
experimental and theoretical results in Fig. 1. Moreover, some
electrons in the area A come out with small final transverse
momenta after scattering with the core (dash lines in Fig. 2(d))
and also form a central interference fringe in Fig. 2(e). This
fringe coincides with the 0th fringe formed by the electrons from
area B (see Fig. 2(c)). It cannot be distinguished in the total
PMD (e. g., see Fig. 1(c)) since the 0th fringe from area B in
this region dominates. It should be mentioned here that electrons
from area C and D only contribute to the background of the
fringes.



It is worthwhile mentioning that the holographic interference
studied in Ref. \cite{Huismans7} arises from the interference
between the EWPs emitted during the same quarter cycle of the
laser field which can be applied to image the sub-cycle dynamics
of the photoelectron. Whereas the newly identified curved
interference structure is induced by the Coulomb-field-driven
transverse backward scattered EWPs that are generated within a
time window of only about 0.06 laser cycle ($\thicksim173$ as for
the 800 nm laser field used here) around the peak of the laser
field. This implies that this novel interference structure can
record electron dynamics on a much shorter time scale.

In the total momentum distribution map, these two interference
structures will coexist and compete with each other. According to
our analysis, the Coulomb potential will become an overwhelming
factor for photoelectrons with small longitudinal velocity when
they come back to the core, leading to this kind of curved
interference fringe. Therefore, the visibility of this structure
in the total momentum distribution is dependent on the proportion
of such photoelectrons that can be strongly affected by the
Coulomb potential. As an estimation, we assume that the return
kinetic energies of such photoelectrons are less than the Coulomb
potential energy at the tunneling exit. Then the interval of the
initial phase around the crest of the laser field for which the
above assumption is valid can be given by
\cite{SupplementalMaterial}
\begin{equation}
\delta\phi\propto\frac{\omega^2}{E_0I_p}\propto\frac{1}{U_pz_0}.\label{5}
\end{equation}
Here $U_p=E_0^2/4\omega^2$ is the ponderomotive energy. According
to Eq. (5), relatively lower intensity and shorter wavelength are
favored by the curved interference fringe which is caused by the
Coulomb-field-driven transverse backward-scattering process.

In fact, this kind of curved interference structure can be clearly
seen in previous experiments of different atoms
\cite{Korneev2012,Marchenko11,Kolesik2014,Richter2015,Marchenko2010}.
It can be found that the curved interference structure in these
experiments, for examples, Fig. 2 in Ref. \cite{Richter2015} and
Fig.21(c) in Ref. \cite{Kolesik2014}, resembles very well the
structure shown in Fig. 2(e), demonstrating that this curved
interference pattern is a``universal" structure for different
atoms in the typical conditions for most current intense field
experiments.

\begin{figure}[h]
\centering
\includegraphics[width=0.30\textwidth,height=0.25\textwidth,angle=0]{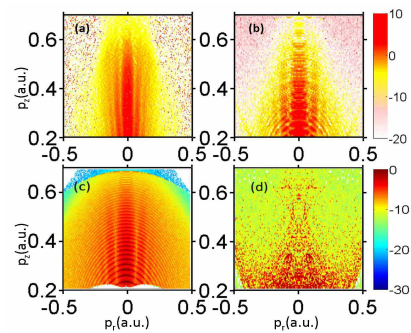}
\caption{(Color online) (a) Experimental two-dimensional
photoelectron momentum spectra of Xe atom in Ref.\cite{Huismans7};
(b) the corresponding GQTMC simulation; (c) the reconstructed
final momentum distributions of electrons in area B. (d) the
reconstructed final momentum distributions of electrons in area A.
}
\end{figure}


In contrast, this curved interference pattern is invisible in the
momentum distribution of Ref. \cite{Huismans7} where a 7-$\mu$m
mid-infrared pulse is used. In this circumstance, though the
ponderomotive energy is relatively small ($U_p=0.118$ a.u.), the
large tunneling exit ($z_0\thickapprox31$ a.u.)
makes the influence of the Coulomb potential very weak. As a
result, the probability of the Coulomb-field-driven transverse
backward-scattering is negligible comparing with the contribution
from area B in Fig. 2(a). Figs. 3(a) and 3(b) show the
experimental result in Ref. \cite{Huismans7} and the corresponding
GQTMC simulation, respectively. Again good agreement is achieved
between the theoretical and experimental results. Both of them
show straight holographic interference fringes and the curved
interference fringes are absent. Figs. 3(c) and 3(d) present the
reconstructed final momentum distributions of electrons in area B
and area A under the same experimental conditions. It can be found
that Fig. 3(c) is almost identical to Fig. 3(b), which means that
electrons only from area B can well reproduce the main
experimental and theoretical PMDs while the electrons from area A
play a negligible role in the total PMD. Moreover, in contrast to
Fig. 2(e), the interference fringes induced by the
Coulomb-field-driven transverse backward-scattering is almost
invisible in Fig. 3(d). This is due to that, in this situation,
the probability of transverse backward scattering is
so low that there are still not enough trajectories in our
simulation (1$\times 10^9$ total trajectories are used) to make
the interference fringes visible in Fig. 3(d).


In summary, a novel curved interference structure is
identified in the experimental and theoretical PMDs of atoms in
intense infrared laser field. A GQTMC method is able to well
reproduce the experimental observation and enable further analysis
of the underlying mechanism of this peculiar structure. We
demonstrated that, different from the well-documented holographic
interference fringes which is attributed to interference between
EWPs generated during a quarter cycle of the laser pulse, the
curved interference structure originates from the interference
among EWPs emitted within an attosecond-timescale window around the
crests of the laser field.
When these electrons are driven back by the laser field to the
core with near-zero longitudinal momenta, they may be pulled back
by the ionic Coulomb potential and be further backward scattered
in the direction perpendicular to the polarization direction. This
scattering may happen at different return times, leading to
different phases of the ejected photoelectrons. The interference
between these electrons results in obvious curved fringes in the
PMD, which can be easily distinguished from the straight radial
holographic interference fringes. Analysis shows that this
interference structure can be observed for different atoms under
the typical conditions of current intense field experiments.
Moreover, this interference pattern can be applied to record
electron dynamics on a time scale of about
100$\sim$200 as.

We are grateful to M. Vrakking and Y. Huismans for providing us
their experimental data. The work was supported by the National
Basic Research Program of China (Grant No. 2013CB922201), the NNSF
of China (Grant Nos. 11374202, 11274220, 11274050, 11334009 and
11425414), Guangdong Natural Science Foundation (Grant No.
2014A030311019), and the Open Fund of the State
Key Laboratory of High Field Laser Physics (SIOM). W. Y.
acknowledges support by the "YangFan" Talent Project of Guangdong
Province.


\begin{thebibliography}{xx}
\bibitem{note} The Coulomb-field-driven transverse backward-scattering may aslo happen at the third or even higher return
times. However, the contribution of these higher return trajectories to the final momentum distribution is negligible according to our calculation.
\bibitem{note2}  Note that the curved interference structure
observed here is different from the structure corresponding to the
outer prong of the fork as discussed in M. M\"{o}ller \emph{et
al.}, Phys. Rev. A \textbf{90}, 023412 (2014) which has nothing to
do with the interference between different electron trajectories
and is not primarily caused by the Coulomb field.
\end{thebibliography}
\end{document}